\documentclass[preprint,12pt]{elsarticle}

\usepackage{subcaption,graphicx}
\usepackage{dcolumn}
\usepackage{bm}
\usepackage{amsfonts}
\usepackage{gensymb}
\usepackage{amsmath}
\usepackage{amssymb}
\usepackage{latexsym}
\usepackage{textcomp}
\usepackage{color}
\usepackage{tikz}
\biboptions{numbers,sort&compress}

\PassOptionsToPackage{normalem}{ulem}
\usepackage{ulem}
\providecolor{added}{rgb}{0,0,1}
\providecolor{deleted}{rgb}{1,0,0}

\journal{Applied Surface Science}

\begin{document}
\def\myfrac#1#2{\frac{\displaystyle #1}{\displaystyle #2}}

\begin{frontmatter}

\title{Optimization of femtosecond laser processing in liquids} 

\author{Jan S. Hoppius}
\ead{hoppius@lat.rub.de}
\address{Applied Laser Technologies, Ruhr-Universit\"at Bochum,
Universit\"atsstra\ss e~150, 44801 Bochum, Germany}

\author{Stella Maragkaki}
\address{Applied Laser Technologies, Ruhr-Universit\"at Bochum,
Universit\"atsstra\ss e~150, 44801 Bochum, Germany}
\address{Institute of Electronic Structure and Laser, Foundation for Research and Technology-Hellas (IESL-FORTH), 71110 Heraklion, Crete, Greece}

\author{Alexander Kanitz}
\address{Applied Laser Technologies, Ruhr-Universit\"at Bochum,
Universit\"atsstra\ss e~150, 44801 Bochum, Germany}

\author{Peter Gregor\v{c}i\v{c}}
\address{Faculty of Mechanical Engineering, University of Ljubljana,
A\v{s}ker\v{c}eva 6, 1000 Ljubljana, Slovenia}

\author{Evgeny L. Gurevich}
\ead{gurevich@lat.rub.de}
\address{Applied Laser Technologies, Ruhr-Universit\"at Bochum,
Universit\"atsstra\ss e~150, 44801 Bochum, Germany}

\date{\today}

\begin{abstract}

In this paper we analyze femtosecond laser processing of metals in liquids searching for optimal conditions for predictable ablation. Incident laser pulses are stretched or compressed, self-focused and scattered on bubbles and on surface waves in the liquid environment. Influence of these effects on the laser intensity distribution on the target surface is discussed and optimal processing parameters are suggested. 

\end{abstract}


\end{frontmatter}


\section{Introduction}

Laser material processing in a liquid environment is potentially advantageous with respect to the commonly used laser ablation in air due to higher processing rates, more effective sample cooling and lower debris left on the surface \cite{KRUUSING1,KRUUSING2}. Further, a surrounding liquid changes the recoil pressure acting on the thin laser-molten layer, which influences formation of the laser-induced nanostructures \cite{Stratakis2009}. Experiments with nanosecond lasers show that the ablation rate of solids in the water confinement regime increases with respect of that in the air \cite{Zhu2001,KangTiSa1ns,Nguyen}. For many applications, like e.g., laser shock peening and generation of nanoparticles, the liquid environment is inevitable. Optimization of the experimental conditions such as pulse overlap and the thickness of the liquid layer increases the productivity of the nanoparticle generation in liquids. For example, maximal production rate of Al$_2$O$_3$ nanoparticles with infrared laser was observed to decrease with the liquid layer thickness \cite{GramScaleProd}. The maximal ablation rate of silicon in water with KrF laser was observed for approximately 1.1\,mm liquid depth, which can be explained by a compromise between light absorption in the liquid and conditions for the plasma formation upon ablation \cite{Zhu}. 
Ultrafast-laser ablation in liquids is a suitable method to fabricate both nanostructures and nanoparticles in a single experiment, with the potential for utilizing both as surface-enhanced Raman scattering (SERS) active elements \cite{Rao}. 

Although liquid environment is obviously advantageous for some aspects of laser ablation, several unwelcome effects appear due to using water or any other liquid as the laser processing medium. Especially for ultrashort laser pulses, which intensity is high enough to trigger nonlinear optical processes in the liquid, the interaction between the latter and the laser light becomes very complex. For example, it was shown, that already for single pulses, the used liquid has a significant impact on the ablation efficiency \cite{Kanitz.2017b}. Furthermore, the laser beam undergoes self-focusing, the pulse duration and spectrum change as well as gas bubbles born upon ablation scatter the following pulses. All these effects lead to unpredictable pulse intensity and spectrum on the target surface and should be controlled to obtain reproducible experimental results. 

Liquid environment is also used in the LIPSS (laser-induced periodic surface structures) formation experiments \cite{Radu,Albu,Barmina,Stella}.
In the frames of plasmonic model of the LIPSS formation, the refractive index of the liquid environment should control the pattern period \cite{Derrien2016}, thus experiments in different liquids could be used for an experimental test of this theory. However the broadening of the spectrum of the incident light in the liquid makes the results difficult for interpretation especially due to the fact that the white light generated in a nonlinear medium can also make the LIPSS \cite{WhiteLight}.

In this paper we analyze the interaction between ultrashort near-infrared laser pulses and liquids experimentally and suggest strategies to minimize these unwelcome effects in order to achieve predictable beam intensity distribution on the sample surface for laser processing.

\section{Experimental Setup}

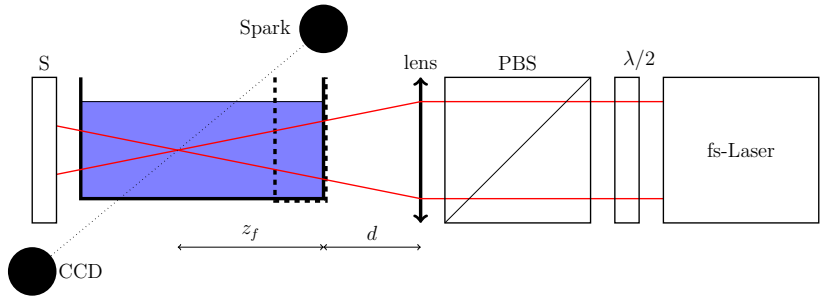
\begin{figure}
    \resizebox{.8\linewidth}{!}{
    \begin{tikzpicture}
  \filldraw[fill=blue!50](0,0)rectangle(5,2);
  \draw[line width=2pt](0,2.5)--(0,0)--(5,0)--(5,2.5);
  \draw[line width=2pt,dashed](4,2.5)--(4,-0.05)--(5.05,-0.05)--(5.05,2.5);
  \draw[line width=2pt,<->](7,-0.5)--(7,2.5)node[above]{lens};
     \draw[thick,red](12,2)--(7,2)--(-0.5,0.5);
     \draw[thick,red](-0.5,1.5)--(7,0)--(12,0);
     \draw[<->](2,-1.0)--(5,-1.0)node[above,midway]{$z_{f}$};
     \draw[<->](7,-1.0)--(5,-1.0)node[above,midway]{$d$};
 \draw[thick](7.5,-0.5)rectangle(10.5,2.5)node[above]{PBS\hspace*{3cm}};
 \draw(7.5,-0.5)--(10.5,2.5);
 \draw[thick](11,-0.5)rectangle(11.5,2.5)node[above]{$\lambda/2$};
 \draw[thick](12,-0.5)rectangle(15.2,2.5)node[midway]{fs-Laser};
     \draw[thick](-1,-0.5)rectangle(-0.5,2.5)node[above]{S\hspace*{0.5cm}};  
        \draw[dotted](-1,-1.5)--(5,3.5);
        \fill(-1,-1.5)circle(0.5cm);
        \fill(5,3.5)circle(0.5cm);
        \node at(0,-1.5){CCD};
        \node at(3.8,3.5){Spark};
   \end{tikzpicture}
   }
 \caption{Experimental setup. Contours of the cuvette filled with the liquid are shown by thick solid lines. Thick dashed line represents the position of a thin cuvette used to analyze white light emission from different slices of the liquid. Distance between the cuvette and the lens is $d$, the distance between the front side of the cuvette and the focal point is $z_f$.}
 \label{Setup}
\end{figure}

In usual experiments on laser processing in liquids, the incident laser beam is focused on the target surface immersed in water or any other liquid either from the top or from the side. The side geometry used in our experiments to study the self-focusing effect is shown in the figure~\ref{Setup}. Femtosecond (fs) laser pulses (\textit{Spitfire Ace} fs-laser produced by \textit{Spectra Physics}, $\lambda=800\,$nm, minimal pulse duration $\tau_p^{min}=35\,$fs) are attenuated by $\lambda /2$ plate and polarizing beam splitter (PBS) and focused into a container by a convex lens (focal distance 8\,cm) from the side. The walls of the cuvettes were one millimeter-thick fused silica glass slides. The container is filled with the liquid under study and contains no target in all experiments, in which only interaction between the laser radiation and the liquid are studied. The geometrical focus point is inside the liquid at the position $z_f$ measured from the front wall of the container. The spectrum of the laser pulses propagating through the liquid is measured by the spectrometer (marked as S in figure~\ref{Setup}) \textit{USB2000+} designed by \textit{Ocean Optics}. The CCD video camera \textit{Point Grey} produced by \textit{FLIR Integrated Imaging Solutions Inc.} and the short-pulse light source (gas discharge spark, signed as Spark in fig.~\ref{Setup}) are placed in the horizontal plane in front and behind the cuvette. The imaging optics (not shown in fig.~\ref{Setup}) homogeneously illuminate the
container and image it onto the CCD chip. The optical axis of the imaging system is perpendicular to the laser beam propagation direction.

In the experiments on bubble formation upon ablation of solid targets immersed in liquids (see section~\ref{SecBubble}), the configuration is closer to the commonly-used laser ablation design: The laser beam impacts the target surface from the top, i.e., through the liquid-air interface. {\it Tangerine} femtosecond laser produced by {\it Amplitude Systems} with $E_p=15\,$\textmu{}J pulses stretched to $\tau_p=1.5\,$ps and repetition rate 1\,kHz was used in these experiments. The beam is focused on a polished copper sample, which is ablated; the position of the laser spot was moved by a Galvo-scanner {\it SCANcube 7} designed by {\it SCANLAB} with the scanning speed of  180\,\textmu{}m/s. The position of the illumination source and the camera are the same as shown in Fig.~\ref{Setup}.

\section{Results and Discussion}
\subsection{Beam self-focusing}

\begin{figure}
 \includegraphics[width=12cm]{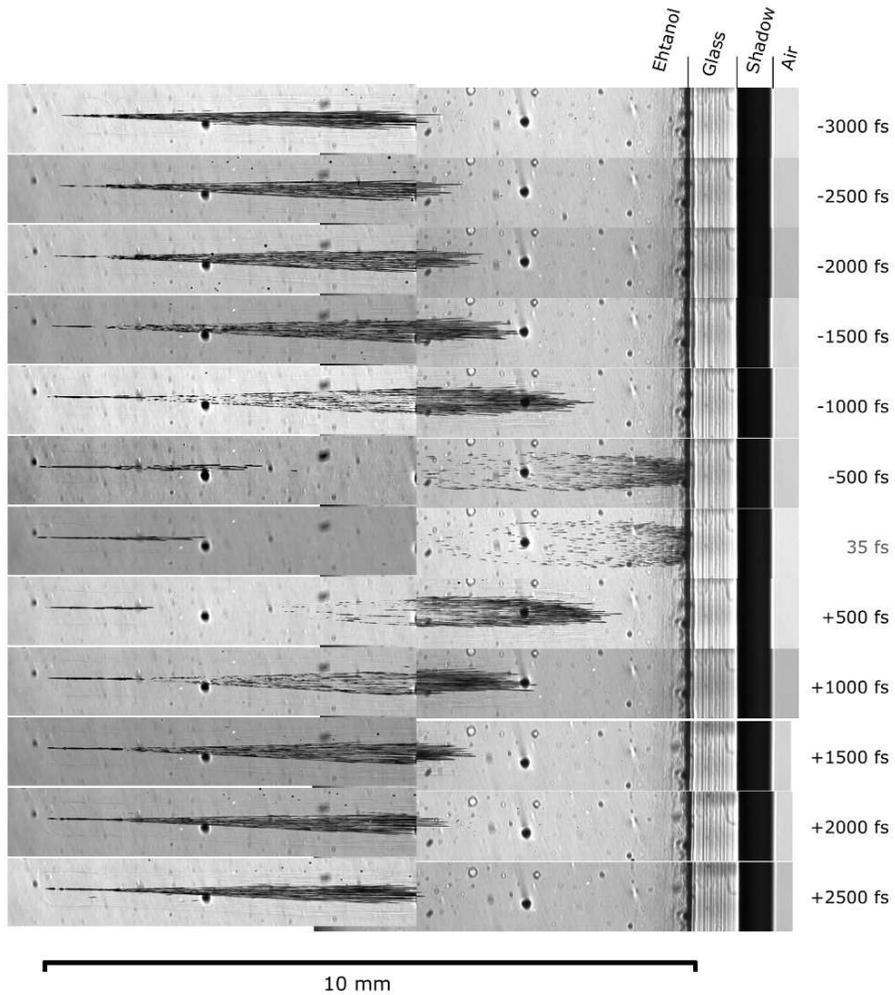}
 \caption{Laser focusing in ethanol, the laser (not shown) is on the right-hand side. Pulse energy $E_p=1\,$mJ, wavelength $\lambda=800\,$nm. Several images corresponding to different pulse durations are shown one above the other (from top to bottom: $\tau_p=-3000\,$fs, $\tau_p=-2500\,$fs, $\tau_p=-2000\,$fs, $\tau_p=-1500\,$fs, $\tau_p=-1000\,$fs, $\tau_p=-500\,$fs, $\tau_p=35\,$fs, $\tau_p=500\,$fs, $\tau_p=1000\,$fs, $\tau_p=1500\,$fs, $\tau_p=-3000\,$fs, $\tau_p=2000\,$fs, $\tau_p=2500\,$fs.). Each image corresponds to single femtosecond laser shot.}
 \label{EthSF}
\end{figure}

A simplified picture of the focusing in liquids is that after the laser beam passes through the air-glass and glass-liquid interfaces it refracts according to the Snell's law and the focus position changes moving slightly further inside the liquid. However, for ultrashort laser pulses this effect is negligible comparing to self-focusing \cite{Marburger.1975}, which shifts the focus in the opposite direction, i.e., closer to the lens. The position of the focus in the container is defined by two lenses: one real lens with focal distance of 8\,cm used in the experimental setup shown in figure~\ref{Setup}, and one virtual lens induced in the liquid with intensity-dependent refractive index $n=n_0+n_2I$ due to the Gaussian beam profile. The optical force of the second lens depends on the parameter $\alpha=\sqrt{P/P_c}$, where $P=E_p/\tau_p$ is the peak power of the pulse, $E_p$ is the pulse energy and $P_c\approx0.15\lambda^2/(n_0n_2)$ - the critical power.
Calculation of $z_f$ \cite{Marburger.1975,Yariv,Couairon.2007} for commonly-used liquids and $\tau_p=35\,$fs laser pulses ($\alpha\sim30000$) reveals that the beam must be self-focused after several millimeters of propagation in the liquid, whereas in the experiments the self-focusing starts either on the liquid interface or already inside the glass wall of the container, see dark filaments in the figure~\ref{EthSF} for $\tau_p=35\,$fs in ethanol.

The onset of the filamentation immediately after the beam enters the medium makes high-power femtosecond lasers nearly useless for precise material processing in liquids.
However, for a given pulse energy $E_p$ the filamentation onset point can be shifted deeper into the liquid by decreasing the parameter $\alpha$, which can be achieved in three ways: (1) by pulse stretching; (2) by changing the focusing conditions; (3) by choosing a liquid with a smaller nonlinear refractive index. 

\subsubsection{Pulse Stretching}

\begin{figure}
 \includegraphics[width=6cm,angle=270]{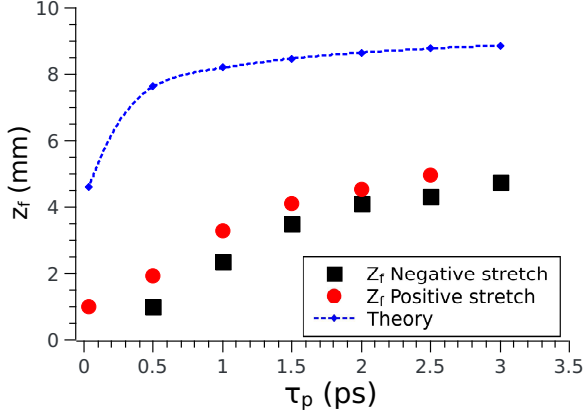}
 \caption{Dependence of the focus position in ethanol $z_f$ on the pulse duration $\tau_p$ for positive and negative pulse stretch measured from the images in figure~\ref{EthSF}. Dashed line is the position of the focus predicted by the equation~\eqref{ZfTheor}. Front wall of the cuvette corresponds to $z_f$ between 0\,mm and 1\,mm, filaments inside the wall cannot be imaged by the optical system.}
 \label{ZfPosition}
\end{figure}

Stretching the pulses by deadjusting the compressor inside the femtosecond laser up to several picoseconds allows to shift the focus several millimeter deeper into the liquid, see figure~\ref{EthSF}. The stretch can be either positive or negative, see figure~\ref{ZfPosition}, a small difference between them can be explained by the group velocity dispersion, which additionally compresses or stretches the laser pulses in the liquid and in the glass. In the figure~\ref{ZfPosition}, $z_f$ is measured as the distance between the front side of the cuvette and the most right dark filament in Fig.~\ref{EthSF}. The thickness of the glass walls are approximately 1 millimeter and the nonlinear refractive index of glass $n_2^{Glass}\approx4\cdot10^{-20}\,m^2/W$ \cite{Sutherland2003} i.e., close to that of water $n_2^{Water}=4.1\cdot10^{-20}\,m^2/W$ and ethanol $n_2^{Eth}=7.7\cdot10^{-20}\,m^2/W$ \cite{Sutherland2003}. Hence we assume that the glass wall doesn't principally change the physical picture of the self-focusing process but the first millimeter of the light propagation in the nonlinear medium (in the glass wall) cannot be imaged by the imaging optics.

It is interesting to compare the observed onset of filamentation (see data points in the figure~\ref{ZfPosition}) with the analytical prediction for $z_f$ \cite{Yariv,Sutherland2003} (see dashed line in the figure~\ref{ZfPosition}) calculated with the equation~\eqref{ZfTheor} in approximation that $\alpha\gg 1$
\begin{equation}
z_f=\frac{R}{\dfrac{R\lambda\sqrt{\alpha -1}}{\pi w_0^2}+1}\approx \frac{R}{\dfrac{R}{\pi w_0^2}\sqrt{\dfrac{E_pn_0n_2}{0.15 \tau_p}}+1}.
\label{ZfTheor}
\end{equation}
Here $R=1\,$cm is the curvature radius of the wave front and $w_0\approx 0.63\,$mm is the laser spot radius; both are estimated at the entrance plane (the interface between the air and the quevette) with the laser beam diameter $D_0=1\,$cm and $d=7\,$cm. One can see that the experimental points have the same trend as the theoretical curve, but the filamentation in the experiments starts considerably earlier than predicted. We assume that this discrepancy can be assigned to the local distortion of the wave front curvature radius $R$, e.g., by scattering on inhomogeneities or by the onset of transverse instability of the beam, which can also be seen in the figure~\ref{EthSF}. We notice that the characteristic transversal length scale of the filaments agrees with the estimation for the period of the fastest growing mode $\Lambda_m=\lambda/\sqrt{2Inn_2}$ \cite{Couairon.2007}, which gives $\Lambda_m\approx 30\,$\textmu{}m for $\tau_p=35\,$fs and $\Lambda_m\approx 300\,$\textmu{}m for $\tau_p=3.5\,$ps.

We notice that for usual femtosecond laser processing of metals, physical processes upon the laser-solid interaction do not change much as long as the pulse duration is below the electron-phonon coupling time, which is of the order of a few picoseconds \cite{Corkum,Stuart,Gamaly}. Hence, femtosecond laser pulses can be stretched if metals are ablated. The situation changes if semiconductors or dielectrics are a processed: here multi-photon and field ionization are important, so the pulse stretching will reduce their efficiency, however, also in this case the ablation threshold weakly depends on the pulse duration for $\tau_p\lesssim 1-10\,$ps \cite{Stuart}.

\subsubsection{Laser Focusing}
Common theory of laser filamentation \cite{Chiao,Marburger.1975} predicts that the light filaments appear as soon as the laser \emph{peak power} reaches the critical value $P_c$ defined by the balance of the light-trapping in a self-induced waveguide and the light diffraction. Hence, any shift of the focal plane with respect to the liquid surface to defocus the beam should not change the self-focusing conditions, since it changes the intensity but not the power. Thus, defocusing should not help avoiding the white light generation and beam filamentation in the liquid.

\begin{figure}
 \centerline{\includegraphics[width=0.5\textwidth]{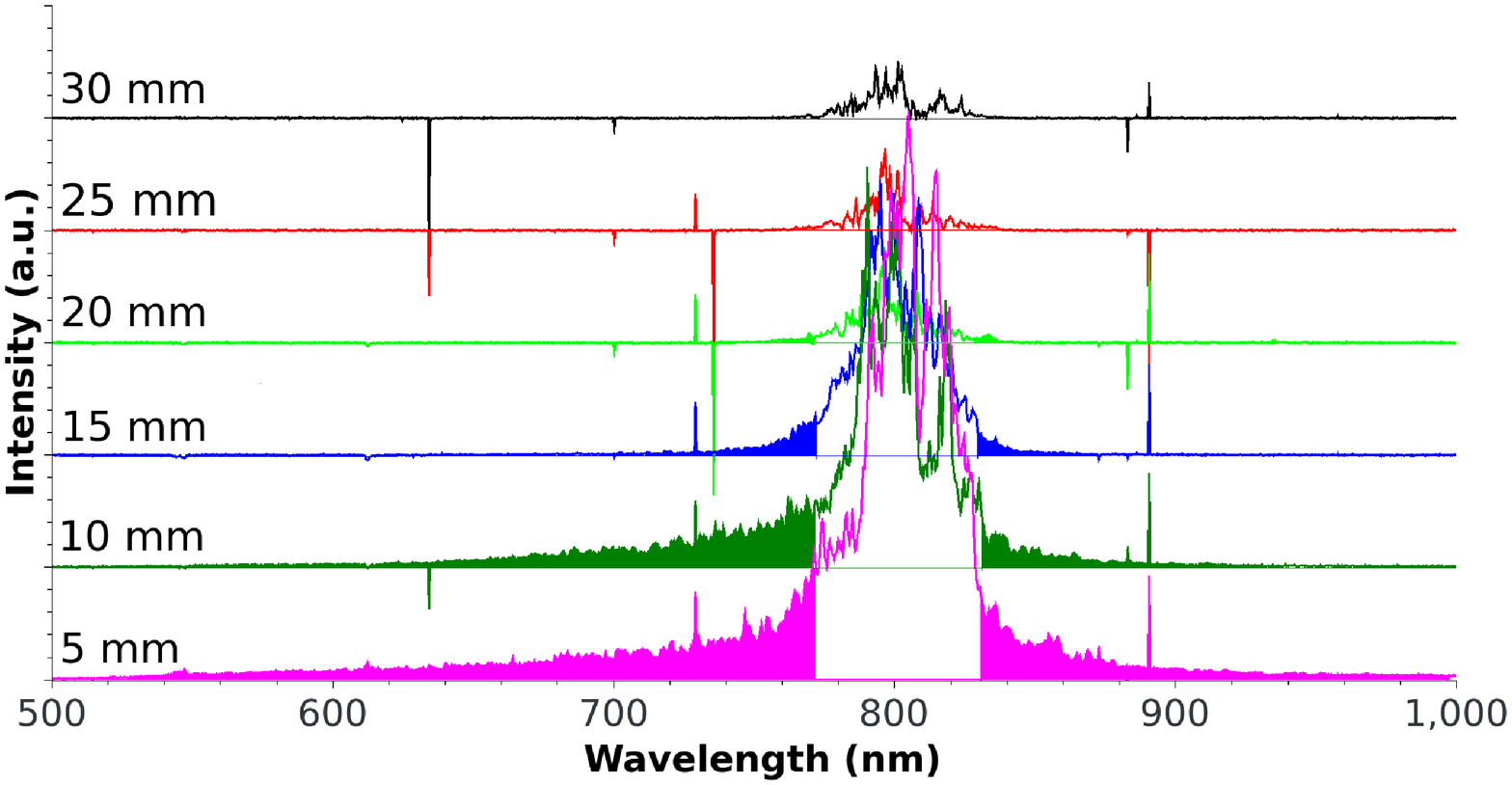}\includegraphics[width=0.5\textwidth]{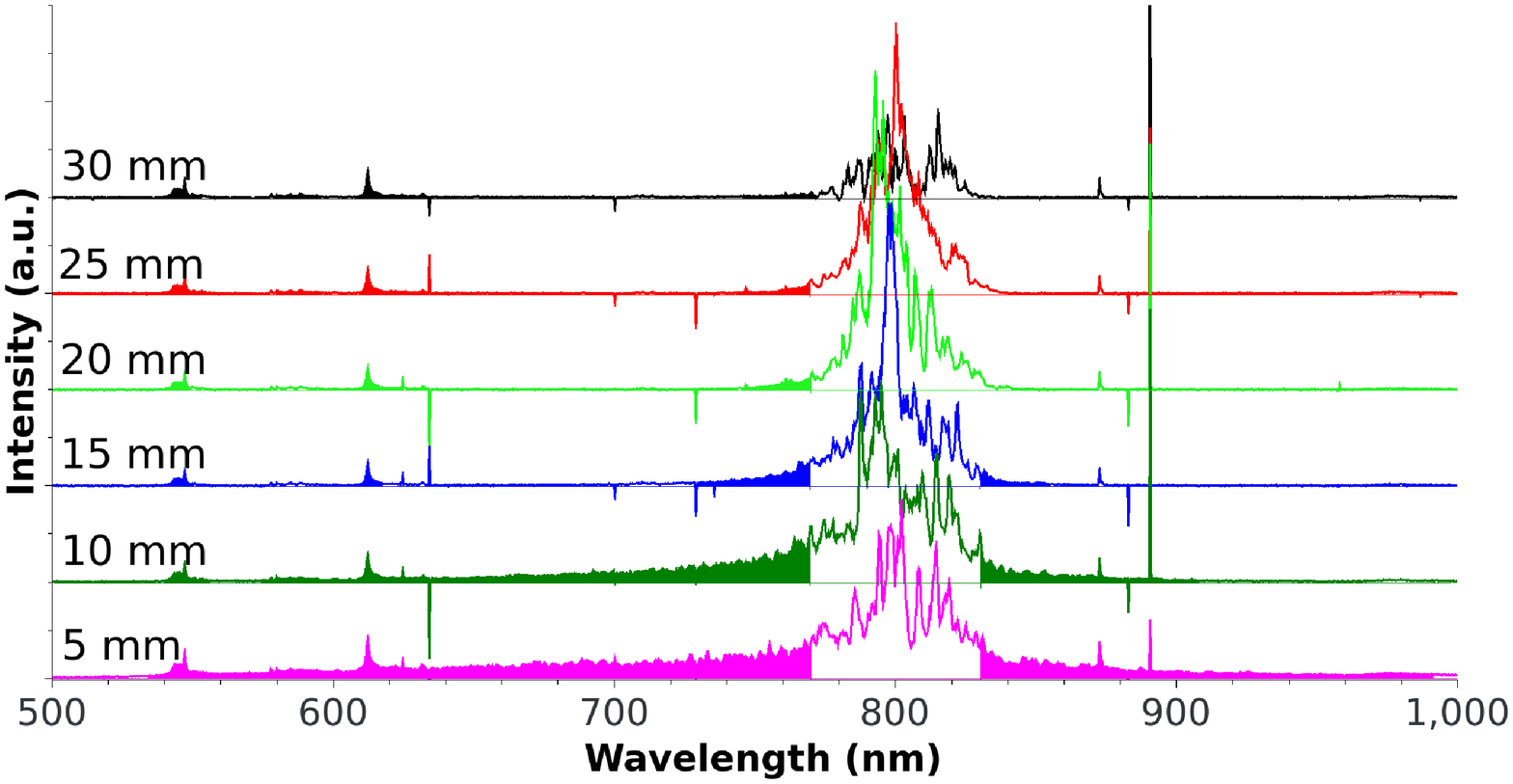}}
 \caption{Spectrum of the white light generated in the 2-mm cuvette placed at different distances $z_f$ from the focus in water (left) and ethanol (right). Spectrum of the incident pulses is Gaussian centered by 800\,nm, line width $\delta\lambda=60\,$nm; the spectrum corresponding to the white light is marked by shaded area below the measured data.}
 \label{WLSpectra}
\end{figure}

We tested this assumption by placing a thin cuvette of $2\,$mm width in the beam between the lens and the geometrical position of the focus in such a way that the focus is in the air behind the cuvette, see dashed line in figure~\ref{Setup}. The distance between the cuvette and the lens $d$ was changed in the experiments, so that the cuvette with the liquid was exposed to the pulses of the same peak power and duration but different intensities. The white light produced in the cuvette at these conditions was analyzed by the spectrometer. Thank to conical emission of the white light \cite{Couairon.2007}, the incident femtosecond laser radiation was blocked by blocking the light coming directly from the laser focus. Thus, the spectra consist of two parts: the scattered incident laser radiation ($\lambda=800\pm30\,$nm) and white light background, which is marked in the figure~\ref{WLSpectra} by the shaded area below the spectrum. It is assumed that the supercontinuum (white light) generation starts simultaneously with the filament formation \cite{Couairon.2007}. Since beam filaments cannot be properly imaged by the optical system in such a thin cuvette, the white light in the spectrum is used as an indicator of the filamentation.

One can see that from the figure~\ref{WLSpectra} that the portion of the white light in the spectrum increases as the cuvette is moved closer to the focus, i.e., by decreasing the laser spot diameter and so increasing the beam intensity. This is not conform with the self-trapping model of the filamentation discussed before, but provides an additional possibility how to avoid the beam filamentation in liquids.

Two following explanations of this effect can be suggested:

1. The self-trapping theory can be corrected by plasma defocusing \cite{Couairon.2007}, which depends on the beam intensity, in this way the size of the laser spot can influence the onset of the filamentation. However the ionization potentials of commonly used liquids are usually larger than 10\,eV, so that for infrared photons seven-photon absorption must be involved to generate plasma. The plasma defocusing should be more pronounced for semi-transparent materials \cite{PhysRevAppl}. 

2. There is another theory \cite{DynamFocus1,DynamFocus2}, in which the filament is described as a dynamic self-focusing effect, in which the focal position is moving following the evolution of the intensity of the laser pulses. The instantaneous position of the focus can be described as the function of the peak power and the spot diameter on the front side of the cuvette \cite{Yariv} (see equation~\eqref{ZfTheor}), hence, the onset of this mechanism can be postponed by defocusing the laser beam.

%

\subsubsection{Liquid Environment}
\begin{figure}
 \includegraphics[width=13cm]{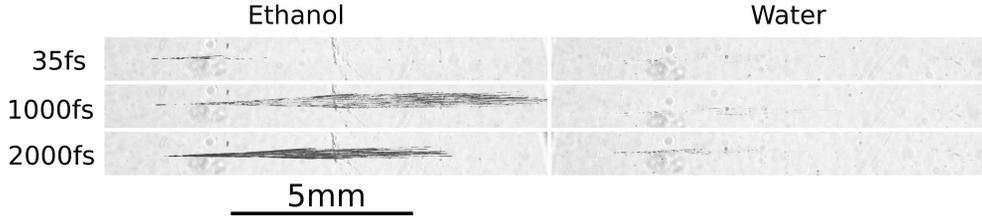}
 \caption{Filamentation pattern observed (a) in ethanol, $n_2^{Eth}=7.7\cdot10^{-20}\,m^2/W$ and (b) in water, $n_2^{Water}=4.1\cdot10^{-20}\,m^2/W$. Characteristics of the laser pulse are: pulse energy $E_p\approx 1\,$mJ/cm$^2$, pulse duration $\tau_p$ is shown left from the images.}
 \label{WaterEt}
\end{figure}

Choosing a liquid with smaller linear and nonlinear refractive indexes, the critical power $P_c$ can be increased. For example the nonlinear refractive index for water is almost two times smaller than that for ethanol \cite{Sutherland2003,NIBBERING1995}. The critical laser peak power, at which filamentation starts, scales with the product ${n_0n_2}$, so one can expect that these nonlinear effects in water will appear at an approximately two-times larger peak power than in ethanol. This estimation is qualitatively confirmed by experimental observations, see figure~\ref{WaterEt}, in which beam filamentation is shown in water and in ethanol for similar processing parameters. 

For pulses, with peak power above the filamentation threshold, the $z_f$ can be increased by decreasing the refractive index according to the equation~\eqref{ZfTheor}. Moreover, if the assumption that the transversal beam instability increases the curvature of the wave front and $R\propto\Lambda_m$ is correct, than $R$ in the equation~\eqref{ZfTheor} should decrease stronger in the liquids with high refractive indexes, which also shifts the onset of filamentation closer to the liquid interface.

Comparing the white light generation in water and in ethanol, see figure~\ref{WLSpectra}, one can see that in water it starts at approximately 20\,mm from the focus of the lens (corresponding to $I\approx 10^{12}$\,W/cm$^2$), whereas in ethanol it starts at approximately 30\,mm (corresponding to $I\approx 0.4\times 10^{12}$\,W/cm$^2$). Thus, white light generation in ethanol starts at a lower incident intensity, but on the other hand, white light generated in water is more intensive. 

\subsection{Bubbles}\label{SecBubble}

Influence of bubbles on the laser ablation in liquids was discussed in the literature \cite{GramScaleProd,BarcikowskiReview,Kalus2017}. The oscillation time $T_0$ of a bubble in an infinite liquid is proportional to its maximum radius $R_{0}$ as $T_0 = 1.83 R_{0} \sqrt{\frac{\rho }{p_0} }$, where $\rho$ stands for the liquid density and $p_0$ is the liquid pressure \cite{Rayleigh,Petkovsek,Brujan}. This means that in case of a bubble with maximum radius of 1 mm in water (density $\rho = 1000$ kg/m$^3$) and at normal conditions ($p_0 = 1$ bar) such bubble expands and collapses within 180 microseconds. However, in case of a laser-induced bubble it disappears usually after several oscillations \cite{XrayNPs,Petkovsek}. Such collapsing bubbles do not importantly influence laser-material interaction in case of femtosecond pulses with the repetition rate in the sub-Megahertz range. However, small gas bubbles may remain in the vicinity of the laser focus for several seconds after laser pulse \cite{Vogel}. They can also stick to the surface and remain there for a very long time. 

These bubbles deteriorate the ablation in the following ways: (1) bubbles generated in the previous laser pulses scatter the following laser pulses; (2) the surface waves on the liquid induced by bubble collapse will additionally focus or defocus the incident laser pulses; (3) if the radius of the bubble is comparable with the thickness of the liquid layer,
these bubbles can break the liquid surface. In this case the liquid can pin to the target surface, so that the following pulses ablate the target in air. This breakup of the liquid surface can only be prevented if the liquid thickness is much larger than the bubble radius.

\begin{figure}
 \centerline{\includegraphics[width=12cm]{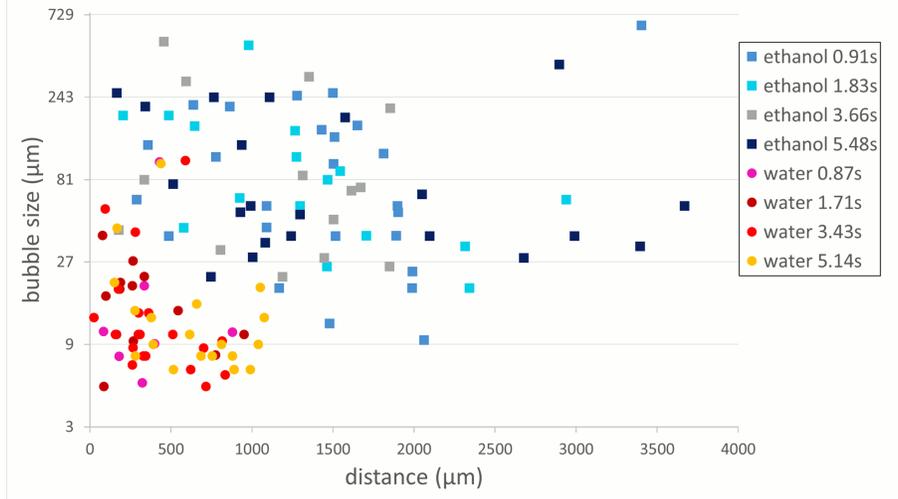}}
 \caption{Bubbles are formed after irradiating a metallic surface in a liquid environment. The graph shows the size of the bubbles and their distance from the laser beam point at different time frames. The squares correspond to irradiation in ethanol and the circles to water. The laser parameters are kept the same (repetition rate $1\,$kHz, $\tau_p=1.5\,$ps, $F=2.7\,J/cm^2$ and scanning speed  180\,\textmu{}m/s).}
 \label{bubbles}
\end{figure}

\begin{figure}
 \centerline{\includegraphics[width=8cm]{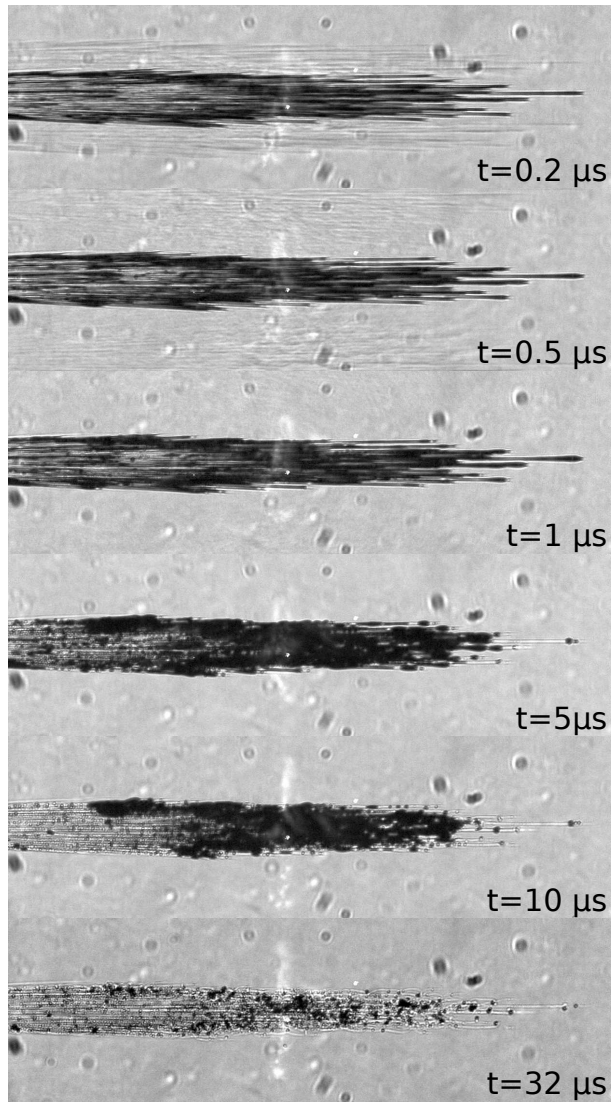}}
 \caption{Breakup of filaments in ethanol into bubbles. Time after the femtosecond laser pulse is shown in each frame.}
 \label{Breakup}
\end{figure}

Statistics of the bubble sizes for water and for ethanol can be found in the figure~\ref{bubbles}. One can see that the pinned bubble size in the ethanol is larger than in water. This difference cannot be explained by the assumption that these bubbles are filled with vapour, which mass $m$ is proportional to the ratio of the pulse energy $E_p$ to the specific energy needed to heat and evaporate the liquid. Indeed, balance of pressures outside (atmospheric pressure $p_0$) and inside the bubble in the equilibrium can be written as $p_a=\frac{m}{M}\frac{RT}{V}$, where $M$ is the molar mass of the vapour molecules, $R=8.31\,$J/(mol K) the gas constant, $T$ - the temperature and $V=\frac{4}{3}\pi R_0^3$ - the bubble volume. Based on this simple model, bubbles produced in our experiments in ethanol and in water must have nearly the same radius, which contradicts to the experimental results, see figure~\ref{bubbles}. This unexpected difference in sizes may indicate that the vapour inside the bubble is partly produced in the filaments, which formation in ethanol is much more intensive, see figure~\ref{WaterEt}. 
This assumption is indirectly supported by the observations of filaments after single laser shots: as one can see in Fig.~\ref{Breakup}, after several microseconds after the laser pulse, the filaments start breaking up into small bubbles.

On the one hand, larger bubbles in ethanol cover larger sample surface and increase the surface screening. On the other hand, these bubbles are removed far away from the focal point on the sample surface by the following laser pulses, see figure~\ref{bubbles}. This can be caused either by lower viscosity of ethanol or by larger drag force, which is proportional to the bubble area. Such a drag force can be induced e.g., by the shock waves generated by the following laser pulses. 

\section{Summary}
Summarizing, practically the self-focusing distance can be shifted deeper inside the liquid by stretching the pulses up to several picoseconds. As soon as the pulse duration is below the electron-phonon coupling time (which is several picoseconds for most of the metals), the physical processes and the ablation rate doesn't change much. We notice also, that the parameter $\alpha$ can be reduced by increasing $P_c\propto\lambda^2$, which can be achieved by choosing a laser with longer wavelength. The optimal thickness of the liquid is defined by a compromise between two effects: on the one hand, it should be less than the self-focusing distance (to avoid self-focusing and supercontinuum generation); on the other hand, it should be larger than the radius of the bubbles generated upon ablation (to avoid surface waves and breakup of the liquid surface). For most of our experiments it corresponds to the liquid height of approximately 4 mm \cite{Alex,Stella}.

\section*{Acknowledgements} Authors acknowledge VIP program of the Research School PLUS, Ruhr-Universit\"at Bochum. 
P. G. acknowledges the financial support from the Slovenian Research 
Agency (research core funding No. P2-0392).



\end{document}